\documentclass[aps, prl,  twocolumn, preprintnumbers]{revtex4-1}

\let\counterwithout\relax

\pdfoutput=1
\usepackage{myoptions}
\usepackage{placeins}
\usepackage{listings}

\usepackage{bbm}
\usepackage[caption=false]{subfig}

\usepackage{chngcntr}

\counterwithout{equation}{section}

\begin{document}

\title{Black holes often saturate entanglement entropy the fastest}

\author{M\'ark Mezei}
\email{mmezei@scgp.stonybrook.edu}
\affiliation{
Simons Center for Geometry and Physics, SUNY, Stony Brook, NY 11794, USA}
\author{Wilke van der Schee}
\email{wilke.van.der.schee@cern.ch}
\affiliation{Theoretical Physics Department, CERN, CH-1211 Gen\`eve 23, Switzerland}

\begin{abstract}

\noindent There is a simple bound on how fast the entanglement entropy of a subregion of a many-body quantum system can saturate in a quench: $t_\text{sat}\geq R/v_B$, where $t_\text{sat}$ is the saturation time,  $R$ the radius of the largest inscribed sphere, and $v_B$ the butterfly velocity characterizing operator growth. By combining analytic and numerical approaches, we show that in systems with a holographic dual, the saturation time is equal to this lower bound for a variety of differently shaped entangling surfaces, implying that the dual black holes saturate the entanglement entropy as fast as possible. This finding adds to the growing list of tasks that black holes are the fastest at. 
We furthermore analyze the complete time evolution of entanglement entropy for large regions with a variety of shapes, yielding more detailed information about the process of thermalization in these systems.

\end{abstract}

\maketitle

\noindent {\bf Introduction:}
The time evolution of entanglement entropy (EE) is an interesting detailed probe of thermalizing many-body systems \cite{bardarson2012unbounded,serbyn2013universal,Hartman:2013qma,Liu:2013iza,Liu:2013qca,Kim:2013bc,kaufman2016quantum}. By causality the EE $S[A(t)]$ of a subregion $A$ can never saturate to its thermal equilibrium value faster than a time $t_{\rm sat} \geq t_\text{LC}$, with $t_\text{LC}$ the time to the tip of the past light cone of region $A$. Geometrically, $t_\text{LC}=R/c$ with 
$R$ the radius of the largest inscribed sphere in $A$ and  $c$ the speed of light \cite{Hartman:2015apr}. Using insight from chaotic operator growth \cite{Shenker:2013pqa,Roberts:2014isa} the bound on the saturation time can be improved to $t_{\rm sat} \geq R/ v_B$ \cite{Mezei:2016wfz}, with $v_B$ the butterfly velocity characterizing the spreading footprint of operators. This improved bound is the main interest of this paper.

\begin{center}
\begin{figure}[t]
\includegraphics[width=0.58\linewidth]{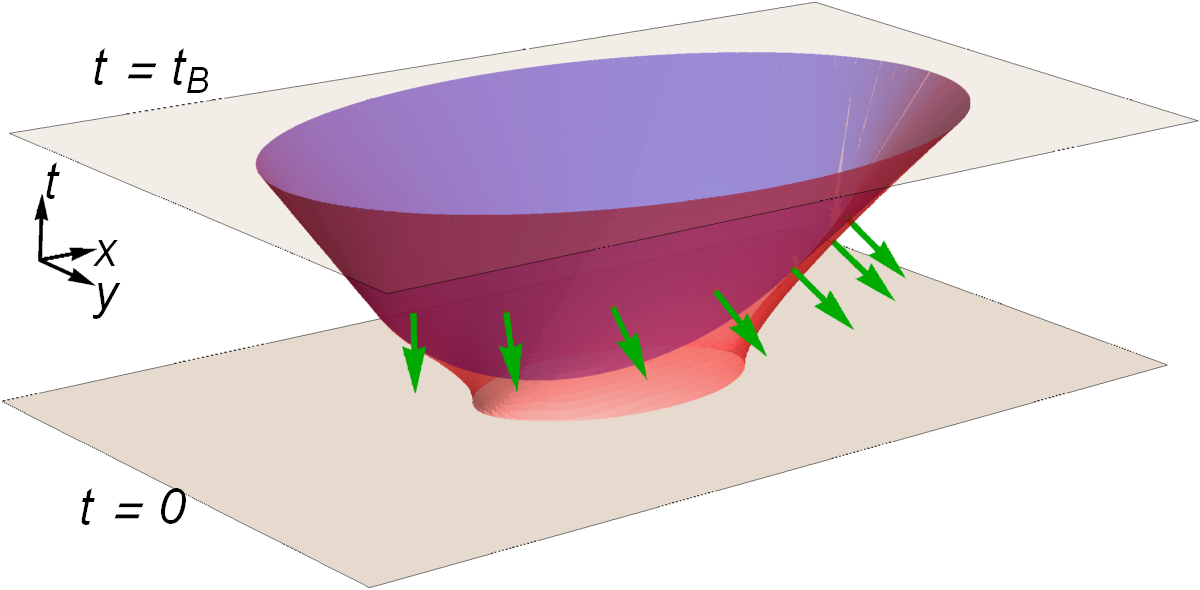}
\includegraphics[width=0.4\linewidth]{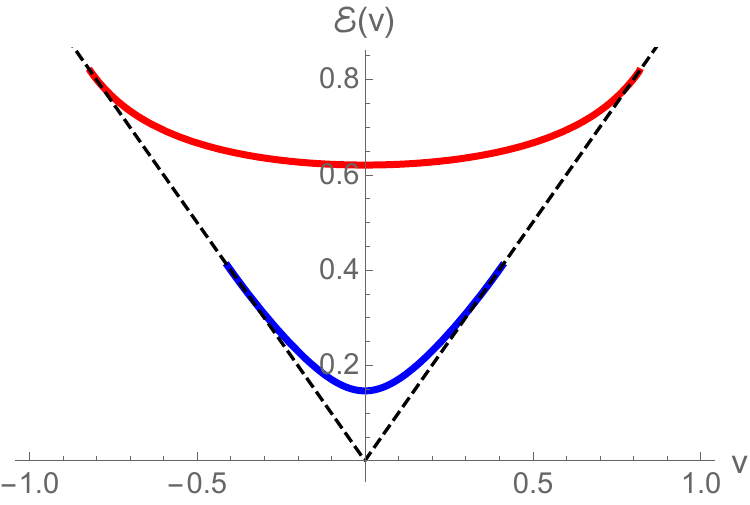}
\caption{(left) A  membrane (red) anchored on an ellipse $A$ shown together with normal vectors (green), the timelike unit vector and the two planes in Minkowski space. The $v_B$  light sheet over the same ellipse (a membrane with $v=v_B$)  is drawn in blue and just touches the plane at $t=0$. 
(right) The membrane tension function of eqn. \eqref{EvSimp} (red) and a charged quench for comparison (blue). The dashed black line is at $45^\circ$. They touch at $v=v_B$. 
\label{fig:membrane}}
\end{figure}
\end{center}

Solving for the time evolution of entanglement entropy is a very challenging problem. Results are available in special solvable examples: two-dimensional CFTs \cite{Calabrese:2005in,Calabrese:2007rg}, free theories \cite{Casini:2015zua,Cotler:2016acd}, random quantum circuits \cite{2015PhRvB..92b4301C,Nahum:2016muy,Nahum:2018pcr}, and holographic gauge theories \cite{Ryu:2006bv,Ryu:2006ef,Hubeny:2007xt,Hartman:2013qma,Liu:2013iza,Liu:2013qca,Mezei:2016wfz,Mezei:2016zxg,Mezei:2019sla}.
Recently, much progress has been made in understanding the process of EE growth in generic chaotic systems
 in a ``hydrodynamic'' limit %
$R,t\gg t_\text{loc}$%
 (with $t/R$ fixed), where  $t_\text{loc}$ is the  local thermalization timescale. For any region, the entropy starts to grow linearly according to the universal law \cite{Calabrese:2005in,Calabrese:2007rg,Hartman:2013qma,Liu:2013iza,Liu:2013qca,Kim:2013bc}:
\es{LinGrowth}{
S[A(t)]&=s_{th} \, v_E\,\text{area}(\p A) t+\dots\,,
}
where $S[A(t)]$ is the vacuum subtracted entanglement entropy of region $A(t)$, $s_{th}$ is the thermal entropy density, and the entanglement velocity $v_E$ is defined by this equation.
In the hydrodynamic limit the growth of EE as a function of time can be described by an effective membrane theory \cite{Nahum:2016muy,Jonay:2018yei,Mezei:2018jco}, which states that 
(the leading piece of) $S[A(t)]$ after a quench at $t=0$ in any chaotic system is computed by the on-shell action of a timelike membrane minimizing the functional
\es{AreaFunctScaled}{
S[A(t)]&=s_\text{th} \int_0^t d\text{area}\ { {\cal E}\le(v\ri)\ov \sqrt{1-v^2}}\,, \qquad v\equiv { (n^\mu \hat t_\mu)\ov \sqrt{1+(n^\mu \hat t_\mu)^2}}\,,
 }
where the timelike membrane stretches between two planes in the $d$-dimensional Minkowski spacetime that the system lives in: it is anchored at $t$ on the upper face on the entangling surface $\p A(t)$ and ends perpendicularly (on an arbitrary shape) on the lower plane at $t=0$. Here $n^\mu$ is its local unit normal ($n^2=1$) and $\hat t^\mu=(1,{\bf 0})$ is the timelike unit vector%
, and the ``velocity'' $v$ is determined by their angle, see Fig.~\ref{fig:membrane}. Since the membrane is timelike it follows that $\abs{v}\leq 1$.  $s_\text{th}$ is the local entropy density and ${\cal E}\le(v\ri)$ is the Lagrangian referred to as the ``angle dependent membrane tension'' that  we elaborate on  below.

The functional was originally presented based on analytic results on random quantum circuits \cite{Nahum:2016muy,Jonay:2018yei}, where the membrane is to be thought of as a coarse grained cut through the tensor network representing the evolving wave function. In holographic gauge theories the EE is given by the area of an extremal surface in the dual spacetime that ends on $\p A(t)$ on the spacetime boundary (that hosts the dual gauge theory) \cite{Ryu:2006bv,Ryu:2006ef,Hubeny:2007xt}. In the hydrodynamic limit, it was shown in  \cite{Mezei:2018jco} that the holographic extra dimension can be integrated out, giving the membrane theory  \eqref{AreaFunctScaled}. In the holographic case the membrane is given by the projection of the extremal surface computing EE in the gravitational description onto the spacetime boundary  along constant infalling time.

These complementary ways of thinking about the membrane add to the intriguing connections between cuts through tensor networks and holographic extremal surfaces pioneered in \cite{Swingle:2009bg,Swingle:2012wq,Hartman:2013qma,Roberts:2014isa,Pastawski:2015qua,Hayden:2016cfa}. %
 The membrane theory can be generalized in many directions, which demonstrates its robustness  \cite{Zhou:2018myl,Bao:2018wwd,Kudler-Flam:2019wtv,Agon:2019qgh,Mezei:2019zyt,Zhou:2019pob}. Since it applies equally to very different systems: holographic gauge theories and random quantum circuits, and since it has all the ingredients to be a universal effective theory of entanglement growth in all chaotic systems, we take the view that it is a universal theory of EE dynamics in much the same way as  hydrodynamics describes the evolution of conserved charge densities; the derivation of the effective membrane theory for general chaotic systems however remains an open challenge.

\noindent{\bf  Specifying ${\cal E}(v)$:} 
Much like the equation of state or transport coefficients in hydrodynamics, the angle dependent membrane tension ${\cal E}(v)$ depends on the theory and conserved charges, but not on other details of the state whose EE dynamics we are studying.   
${\cal E}\le(v\ri)$ obeys the following general constraints:
it is an even function of $v$, monotonically increasing for $v>0$, convex, interpolates between ${\cal E}\le(0\ri)=v_E$ and ${\cal E}\le(v_B\ri)=v_B$, and is tangent to the $45^\circ$ line at $v_B$, ${\cal E}'\le(v_B\ri)=1$. %

In random quantum circuits ${\cal E}(v)$ depends on the structure of the circuit: in one simple 1+1D example with a large on-site Hilbert space, ${\cal E}(v)=\tfrac{1}{2}(1+v^2)$ \cite{Nahum:2016muy}, and no results are known for higher dimensions. In holography, ${\cal E}(v)$ repackages the dual spacetime geometry. A field theory quench is dual to a spacetime in which a  black hole forms from collapse.  ${\cal E}(v)$  only depends on the static region of spacetime that contains the equilibrium black hole that the out of equilibrium initial state settles to, and in particular does not depend on the details of the quench. This dual metric can be written as:
\es{BHmetric}{
ds^2&={1\ov z^2}\le[-a(z) dt^2-2dt dz+d\vec{x}^2\ri]\,,\\
a(z)&=1 - M z^d+ Q^2 z^{2(d-1)} \,,
}
where $(t,\vec{x})$ are the field theory coordinate and $z$ is the holographic extra dimension, and we chose a family of charged black holes as an example with $M$ the mass and $Q$ the charge, which map to the energy and charge density of the state in the field theory. 
 Then ${\cal E}(v)$ is given parametrically by the formula:
\es{Evparametric}{
\le\{v(z),{\cal E}(v(z))\ri\}=\le\{\sqrt{a(z)-{za'(z)\ov 2(d-1)}}\,,\sqrt{-a'(z)\ov 2(d-1) z^{2d-3}}\ri\}\,,
}
where $z\in[0,z_*]$ and $z_*$ is the value for which $v(z_*)=0$ \cite{Mezei:2018jco}. In this paper we will study in detail the case $d=4,\, Q=0$, describing charge neutral quenches in 4D holographic field theories, for which \eqref{Evparametric} evaluates to:
\es{EvSimp}{
{\cal E}\le(v\ri)&={v_E\ov (1-v^2)^{1/4}}\,,\quad v_E={\sqrt2\ov 3^{3/4}}\,, \quad v_B=\sqrt{2\ov 3}\,,
}
 see Fig.~\ref{fig:membrane} (right).
%
\footnote{In 2D every subregion is an interval, and the entropy grows linearly with time with the rate $v_E$. In 3D even spheres do not saturate the saturation time bound \cite{Mezei:2016zxg,Mezei:2018jco}, hence we chose to study the 4D case.}
  On occasion we will present results for charged black holes, whose dynamics is expected to be slower due to the decrease of the size of the available Hilbert space \cite{Leichenauer:2014nxa}. Our methods and (most of our) results straightforwardly extend to other ${\cal E}\le(v\ri)$'s and hence conjecturally to any chaotic system. While currently the only higher-dimensional examples for  ${\cal E}\le(v\ri)$ come from holography, once  new analytic or numerical ${\cal E}\le(v\ri)$ functions become known for non-holographic systems, it will be very interesting to revisit our results for them.
  
   In the following we compute the time evolution of EE in the membrane theory \eqref{AreaFunctScaled} with ${\cal E}(v)$ given in \eqref{EvSimp} (and some charged generalizations) for a variety of entangling regions, which compute the EE in holographic field theories in the hydrodynamic regime. Besides determining the full time evolution, we analyze the saturation time in detail.

%
%
%
%


%
%
%

%

%

%
%
%
%
%
%
%
%

\noindent{\bf Analytic results:} We briefly review the analytic solution of the membrane theory for symmetric shapes. For $A$ a strip, at early times by symmetry the membrane is a straight plane stretching between the $t=0$ and $t$ time slices. Evaluating \eqref{AreaFunctScaled} for this membrane gives linear growth with slope $s_\text{th} v_E\, \text{area}[\p A]$ until saturation, consistent with \eqref{LinGrowth}. The saturation time for a strip of width $2R$ is hence $t_\text{sat}=R/v_E \equiv t_E$.
For spherical $A$, the action reduces to
\es{AreaFunctSphere}{
S[A(t)]&=4\pi \, s_\text{th} \int_0^t ds\ \rho^2(s)\,{ {\cal E}\le(\dot{\rho}(s)\ri)}\,,
 }
where $s$ is the time coordinate $0\leq s \leq t$ and we describe the membrane with a radial coordinate $\rho(s)$ (hence $v=\dot{\rho}(s)$). Minimizing \eqref{AreaFunctSphere} is a one dimensional classical mechanics problem, and can be straightforwardly solved using energy conservation.
$S[A(t)]$ is shown in Fig.~\ref{fig:manystadia} (as the $r=1$ curve) for the membrane tension corresponding to the neutral black hole \eqref{EvSimp}.  The curve ends at $t_B$: the corresponding membrane has $v(s)=v_B$ and hence is a $v_B$ light sheet.
The problem   for a cylinder subregion is  solved by replacing $4\pi \, \rho^2(s) \to 2\pi L \, \rho(s)$ in \eqref{AreaFunctSphere}. Remarkably, 
the cylinder also saturates at $t_\text{sat}=t_B$, see Fig.~\ref{fig:manystadia} (as the $r=\infty$ curve). Below we will find that shapes that interpolate between the sphere and cylinder saturate EE at $t_B$, while those that interpolate between the sphere and the strip have $t_B\leq t_\text{sat}<t_E$. 

We have also analyzed charged quenches whose EE dynamics is governed by ${\cal E}(v)$ computed from 
 \eqref{Evparametric}. We find three regimes as a function of $q\equiv Q/M^{3/4}$: for $0\leq q \leq 0.38$ both the sphere and the cylinder (and we expect that all the shapes that interpolate between them) have $t_\text{sat}=t_B$; for  $0.38 \leq q \leq 0.61$ the sphere has  $t_\text{sat}=t_B$, while the cylinder has $t_\text{sat}>t_B$  (and hence among the shapes that interpolate between them, there should be an open set with $t_\text{sat}=t_B$); while for $0.61 \leq q $ all shapes have $t_\text{sat}>t_B$.  We present a detailed discussion of  $S[A(t)]$ for a spherical region from this last regime of $q$, for $q=0.62$ in the Supplemental Material (SM). 

\begin{figure*}[!htbp]
 \includegraphics[width=0.98\textwidth]{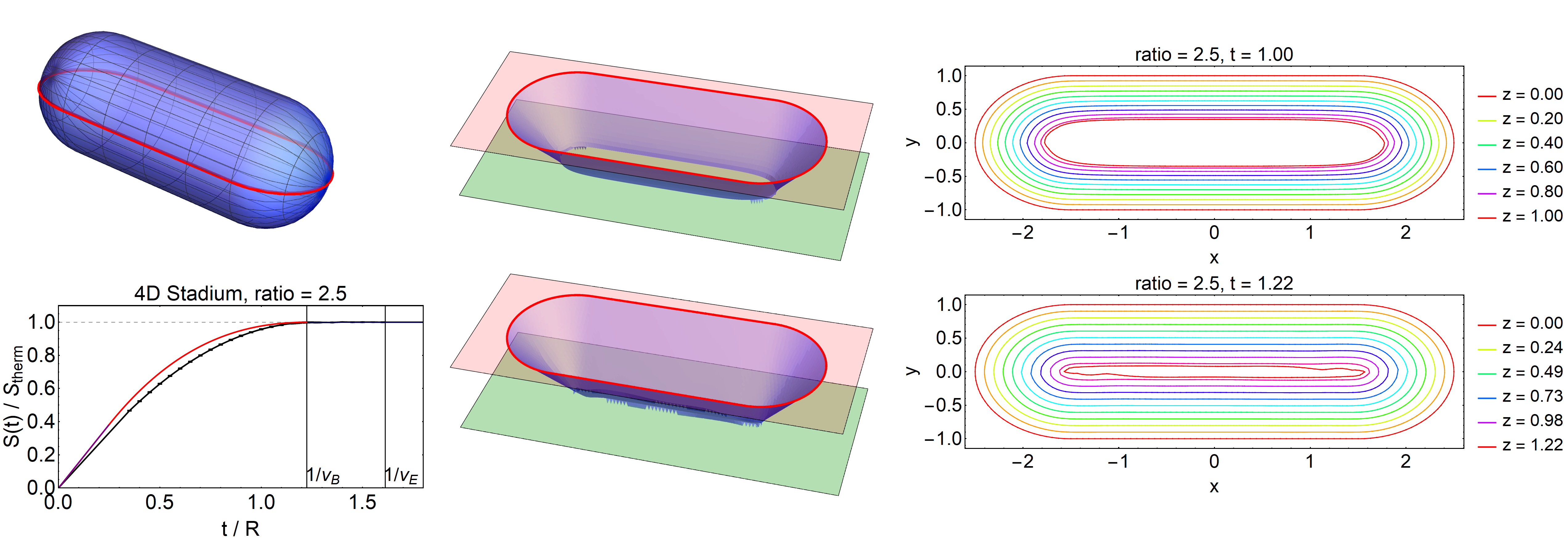}%
\caption{Time evolution of the entanglement entropy of rotated stadia with short to long ratio of $2.5$, including two detailed illustrations of the relevant membranes as well as the analytic bound (red, eqn.~\ref{Sbound}). The largest inscribed sphere has radius $R=1$. Curiously even for this non-trivial shape the entanglement entropy saturates as fast as possible, at the butterfly time $t_B$.
}
\label{fig:examples}
\end{figure*}

\begin{figure*}[!htbp]
 \includegraphics[width=0.98\textwidth]{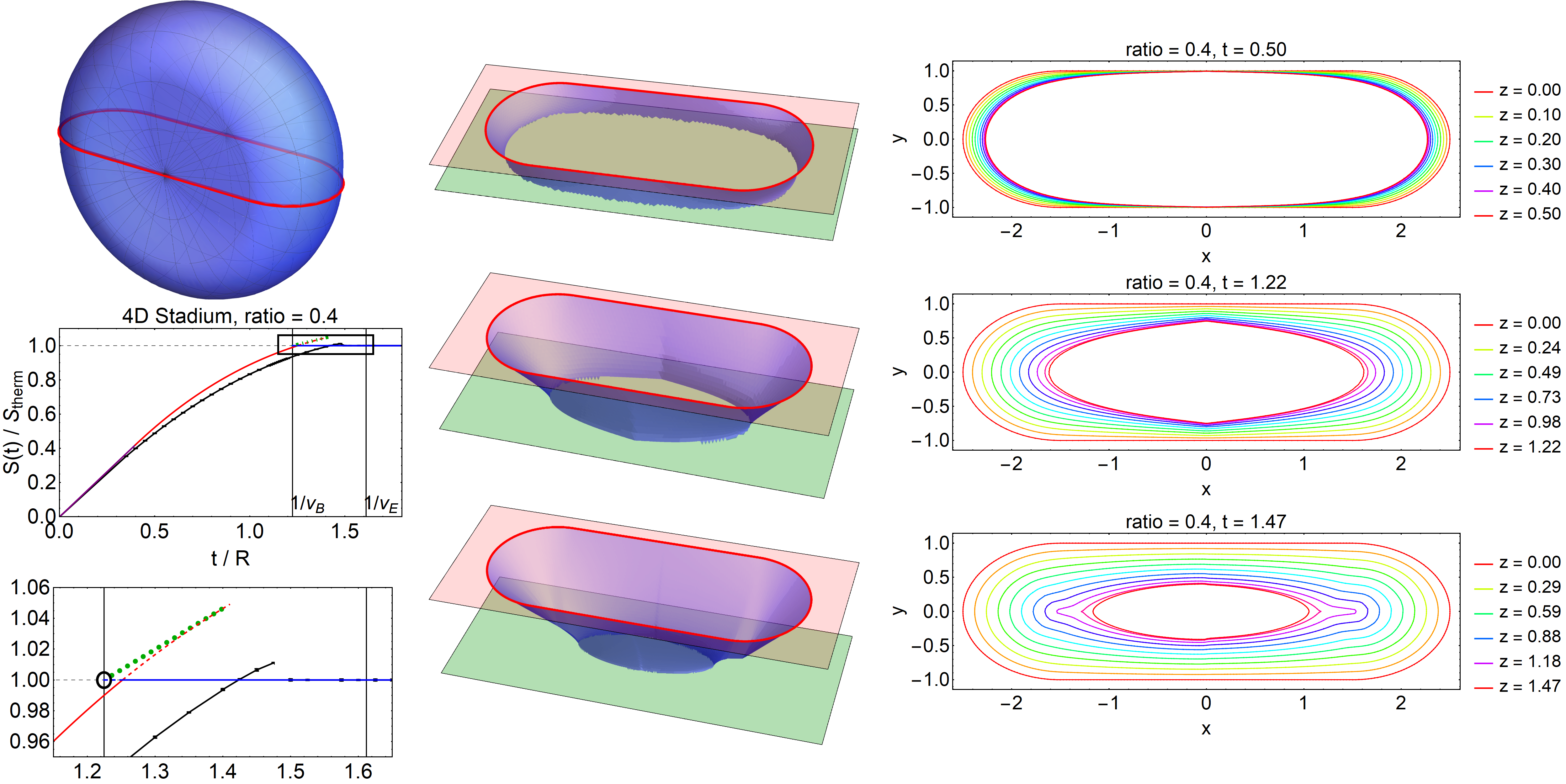}%
\caption{Time evolution of the entanglement entropy of rotated stadia with short to long ratio of $0.4$. The time evolution obeys the analytic bound of eqn.~\ref{Sbound}, plotted by a red solid line when they are minimal and dashed where they become non-minimal. The $v_B$ light sheet at $t_B$ is indicated by a black circle, which lies at the end of a dashed green line representing a disfavored branch of variational membranes.
}
\label{fig:examplesR}
\end{figure*}

For more general shapes the minimal membrane will be solved numerically in the next section, but it is possible to  obtain rather constraining analytic upper bounds on the entropy \cite{Mezei:2016wfz}. In the framework of the membrane theory \eqref{AreaFunctScaled} this bound corresponds to considering a variational surface consisting of two parts joined at $t=t'$: a ``light sheet" part of slope $v_B$ and a vertical tube with $\mathcal{E}(v=0)=v_E$. We get a tight upper bound by minimizing in $t'$ the total membrane action:
\es{Sbound}{
S_\text{max}[A(t)]=&s_\text{th}\min_{0\leq t' \leq \min(t,t_B)}\le[\le(\vol\le(A\ri)-\vol\le(A'\ri)\ri)\ri.\\
&\le.+v_E\,\text{area}\le(A'\ri)\, t' \ri]\,,
}
where $A'(t')$ is the past butterfly domain of dependence of $A(t)$ at time $t'$, i.e. the set of points in region $A$ that are farther from $\p A$ than $v_B(t-t')$. We show $S_\text{max}[A(t)]$ on Fig.~
\ref{fig:examplesR} together with the numerical results for $S[A(t)]$. The bounds are very close to the actual results 
 (see also \cite{Mezei:2016wfz} for similar results for spherical regions).

\noindent{\bf Numerical results:}
In general the minimization of the action \eqref{AreaFunctScaled} cannot be solved analytically, but it is possible to start with some initial surface and gradually relax this surface to a (local) minimal solution. For this we used the Surface Evolver \cite{brakke1992surface}, which uses a triangulation of the surface to minimize some energy functional. We implemented this in 3D for ellipses and stadia (consisting of two half-circles connected by lines) having a ratio $r$ between the long and short axes. In 4D we kept one rotational symmetry by rotating these surface about the long axis (Fig.~\ref{fig:examples}) or short axis (Fig.~\ref{fig:examplesR}). The numerical implementation is included in the SM.

\begin{center}
\begin{figure}[!htbp]
\includegraphics[width=0.85\linewidth]{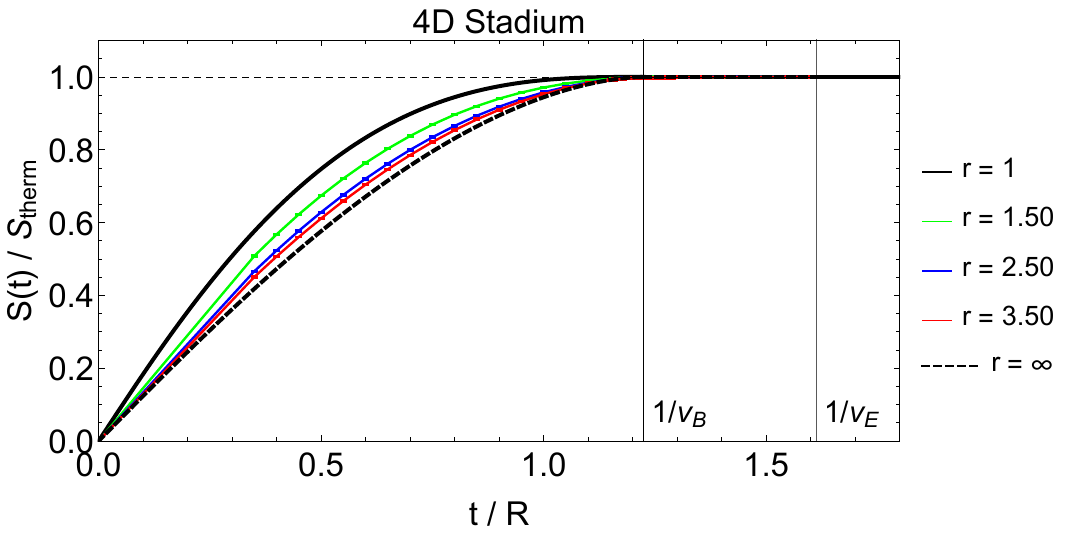}
\includegraphics[width=0.85\linewidth]{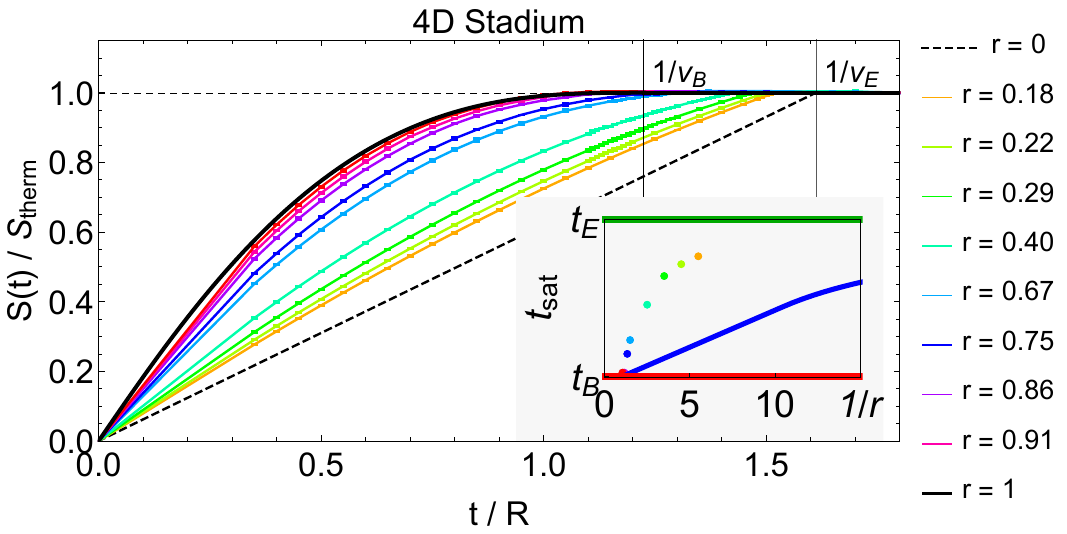}
\caption{ 
See Fig.~\ref{fig:examples} for illustrations of the rotated stadia. Here we show for several shapes the entanglement entropy. The analytically solvable sphere, cylinder and strip are shown in black and correspond to ratios $1$, $0$ and $\infty$ respectively. The inset shows the saturation time (colored dots) for rotated stadia, together with the analytic lower bound (solid blue) coming from Eqn.~\ref{Sbound} (see the red line in Fig.~\ref{fig:examplesR} for an example with $r=0.4$).
\label{fig:manystadia}}
\end{figure}
\end{center}

We now briefly describe some general features of $S[A(t)]$ and the membranes computing it, shown in Figs.~\ref{fig:examples}-\ref{fig:manystadia}. 
 In the early time regime captured by \eqref{LinGrowth}, the membrane has $v\approx 0$, i.e. it is a tube stretching between the upper and lower ends of the spacetime slab (see see Fig.~\ref{fig:examplesR} at $t=0.5$). After this early time regime, $S[A(t)]$ starts to curve, and smoothly saturates to the thermal value $S[A(t)]=s_\text{th}\, \vol(A)$ at some saturation time $t_\text{sat}$. At intermediate times the membrane generically forms cusps, most visible on Fig.~\ref{fig:examplesR}. Since the action depends on $v$ but not its derivatives, the jump of $v$ at the cusp does not cause a divergence in the action.

\noindent{\bf Saturation time:}
A remarkable feature of Figs.~\ref{fig:examples} and~\ref{fig:manystadia}  is that EE saturates at the butterfly time $t_B=R/v_B$, where $R$ is the radius of the largest inscribed sphere inside the entangling surface.  We found the same feature in prolate ellipsoids; the numerical results are included in the SM. Independent of the validity of the membrane theory, \cite{Mezei:2016wfz} gave an argument that the bound $t_\text{sat}\geq t_B$ holds in any many-body system. The saturation of this bound is our most important result and allows us to conclude that {\it neutral (and moderately charged) black holes often saturate entanglement entropy the fastest}, where ``often'' indicates a large class of shapes. 

We now explain the ``often''  qualifier of this statement. 
It is not true that for all shapes the saturation time is as fast as possible. For the analytically solvable case of the strip, we get $t_\text{sat}=t_E>t_B$. (Recall that $t_E=R/v_E$.) For squashed shapes, the stadium of Fig.~\ref{fig:examplesR} and also other stadia on Fig.~\ref{fig:manystadia}, as well as oblate ellipsoids (see the SM), we find $t_\text{sat}>t_B$. In the case of the stadia, this can be proven analytically: %
 we  show that even $S_\text{max}[A(t)]$ saturates later than  $t_B$, and hence so does $S[A(t)]$ (see Fig.~\ref{fig:examplesR} for an example).  In the inset of Fig.~\ref{fig:manystadia} we show the analytic lower bound on the saturation time that interpolates between $t_B$ and $t_E$ together with the numerical results. That for $r\to 0$ we get $t_\text{sat}=t_E$ is expected, since in this limit the squashed stadium becomes a strip. We note that in the membrane theory one can prove a simple upper bound on the saturation time, $t_\text{sat}\leq t_E$ \cite{Mezei:2018jco}, so the family of shapes that we consider realize all possible saturation times.

It would be bold to conclude that $t_\text{sat}=t_B$ for a large family of shapes based solely on numerical data. We now make our case stronger by presenting a semi-analytic argument to this effect.
Interestingly, we can establish analytically that at $t=t_B$ the $v_B$ light sheet over the entangling surface is a (locally) minimal  surface \cite{Mezei:2018jco}. (This is also a surface that is contained in the set of variational surfaces in \eqref{Sbound} for $t'=0$.) In some cases it is also the global minimum, and we have $t_\text{sat}=t_B$, see Fig.~\ref{fig:examples}. In other cases, as shown on Fig~\ref{fig:examplesR}, lower left corner, we see that there is a variational surface with $t'>0$ 
that gives lower entropy (red line at $t=t_B$), even though the numerical membrane gives even lower entropy (black line). At some time $t_{\rm sat, var}>t_B$ the variational surface gives a larger entropy than the thermal value, after which we continue the (disfavored) variational estimate as dashed line. It is this transition that gives a lower bound on the actual saturation time, as shown in the inset of Fig.~\ref{fig:manystadia}. At an even later time it is not possible to connect the two surfaces with a regular minimal variational membrane; we indicated the smooth transition from that point to the $v_B$ lightsheet (black circle) by a dashed green line on Fig~\ref{fig:examplesR}.

 Numerically we were only able to follow the favored
branch, but we believe that the disfavored branch also exists. 
These two branches are the only ones that give membranes that connect the two ends of the spacetime slab. On top of these, the membrane theory also allows for a horizontal membrane that gives $S[A(t)]=s_\text{th}\vol(A)$. This is the horizontal line on all figures, colored blue on Fig~\ref{fig:examplesR}. For $t>t_\text{sat}$ it is the minimal membrane.

The existence of multiple branches of minimal membranes is the mechanism by which the $v_B$ light sheet fails to be the relevant membrane at  $t=t_B$.  Another piece of evidence for this mechanism comes from the analytic study of spherical entangling surfaces for membrane tensions ${\cal E}(v)$ that result from a charged quench in holographic theories, where a favored and disfavored branch was seen in \cite{Mezei:2016zxg} (for completeness included in the SM). From understanding the mechanism for how we get $t_\text{sat}>t_B$ it follows that we only have to understand if the two branches exist or not. In the latter case, we get $t_\text{sat}=t_B$. 
In this paper we decide this question numerically.

We also note that depending on which situation we are in, the slope of the curve, ${d\ov dt}S[A(t)]\vert_{t=t_\text{sat}}=\text{finite}$ for the case $t_\text{sat}>t_B$, and zero when $t_\text{sat}=t_B$ (see Fig.~\ref{fig:examplesR}). 

\noindent{\bf Discussion:} In this paper we studied the entanglement entropy $S[A(t)]$ in the hydrodynamic regime, of large regions at late times, in the membrane effective theory proposed to capture the dynamics of entanglement in all chaotic systems. By focusing on the membrane tension function ${\cal E}(v)$ given in \eqref{EvSimp}, we specialized to the case of 4D holographic gauge theories and neutral quenches, but all our methods generalize to any theory, once ${\cal E}(v)$ is provided as an input. We also derived some results for charged quenches in the same theories.

One important  problem we leave for the future is how to determine whether $t_\text{sat}=t_B$ or larger analytically without having to solve for the minimal membrane numerically. This may be possible by perturbing around the $v_B$ light sheet membrane to decide if it is on the dominant or disfavored branch, as in 
 Fig.~\ref{fig:examples} or  Fig.~\ref{fig:examplesR} respectively.

While in the case of holography, we have a way of computing not just the leading piece in the hydrodynamic limit,  but the exact entropy (using the holographic surface extremization prescription \cite{Ryu:2006bv,Ryu:2006ef,Hubeny:2007xt}), even in this context there are many advantages of using the membrane theory: the simplified description allows for the identification of features that survive the hydrodynamic limit, enables the understanding of near saturation behavior, and the numerical solution of the relatively simple-looking surface extremization problem is prohibitive in the presence of a large scale separation $R,t\gg \beta$. In generic chaotic theories the determination of the exact entropy is out of reach, but one may hope that in the future we will learn how to determine  ${\cal E}(v)$ from other data characterizing the theory. One hint that this may be possible is that we can determine the special point ${\cal E}(v_B)=v_B$ by studying out-of-time-order correlators. Another way to exploit the power of the membrane theory is to determine  ${\cal E}(v)$ from determining $S[A(t)]$ e.g. for a sphere in numerics (see \cite{Jonay:2018yei} for a related numerical determination of ${\cal E}(v)$) or in the future in experiments (see \cite{kaufman2016quantum} for early experimental work), which then yields infinitely many new predictions for other shapes.

In the future, we can use the numerical methods developed here to study the entanglement entropy in inhomogeneous quench setups, where the membrane theory is coupled to the hydrodynamic degrees of freedom  \cite{Mezei:2019zyt}. One fascinating application would be to understand whether signatures of turbulent fluid flows show up in entanglement entropy.

Our most interesting finding is that moderately charged black holes produce entropy dynamics with saturation time $t_\text{sat}=t_B$ for elongated shapes that interpolate between a sphere and the cylinder, which is the fastest possible allowed by quantum mechanics \cite{Mezei:2016wfz}. This adds to the list of things black holes excel at: they are also fastest scramblers \cite{Hayden:2007cs,Sekino:2008he,Shenker:2013pqa}, have Planckian equilibration time \cite{Chesler:2008hg,Heller:2011ju}, and are conjectured to have the lowest sheer viscosity to entropy density ratio in nature \cite{Kovtun:2003wp}.
 
To highlight how efficient black holes are at saturating entropy, we  contrast their chaotic dynamics to those of free field theories (in Gaussian states), whose EE dynamics is expected to be modeled by the quasiparticle theory \cite{Calabrese:2005in,Calabrese:2007rg,Casini:2015zua,Cotler:2016acd}. As shown in \cite{Casini:2015zua}, this model gives $t_\text{sat}=d_\text{max}/2$, where $d_\text{max}$ is the largest distance between two points of the subregion. The intuition behind this result is simple: this is the time when one member of any quasiparticle pair  free streaming at the speed of light has left the subregion $A$.  Since we have to associate $v_B=1$ to these systems, only spheres can have  $t_\text{sat}=t_B$, and all elongated shapes give longer saturation times $t_\text{sat}>t_B$ with $t_\text{sat}=\infty$ for the cylinder.

\noindent{\bf Acknowledgments:} We thank Julio Virrueta and Amos Yarom for useful discussions.  
MM thanks the KITP for its hospitality during the final stages of this work, where his research was supported in part by the National Science Foundation under Grant No.~NSF PHY-1748958.

\bibliography{sattime_refs}

\begin{thebibliography}{43}%
\makeatletter
\providecommand \@ifxundefined [1]{%
 \@ifx{#1\undefined}
}%
\providecommand \@ifnum [1]{%
 \ifnum #1\expandafter \@firstoftwo
 \else \expandafter \@secondoftwo
 \fi
}%
\providecommand \@ifx [1]{%
 \ifx #1\expandafter \@firstoftwo
 \else \expandafter \@secondoftwo
 \fi
}%
\providecommand \natexlab [1]{#1}%
\providecommand \enquote  [1]{``#1''}%
\providecommand \bibnamefont  [1]{#1}%
\providecommand \bibfnamefont [1]{#1}%
\providecommand \citenamefont [1]{#1}%
\providecommand \href@noop [0]{\@secondoftwo}%
\providecommand \href [0]{\begingroup \@sanitize@url \@href}%
\providecommand \@href[1]{\@@startlink{#1}\@@href}%
\providecommand \@@href[1]{\endgroup#1\@@endlink}%
\providecommand \@sanitize@url [0]{\catcode `\\12\catcode `\$12\catcode
  `\&12\catcode `\#12\catcode `\^12\catcode `\_12\catcode `\%12\relax}%
\providecommand \@@startlink[1]{}%
\providecommand \@@endlink[0]{}%
\providecommand \url  [0]{\begingroup\@sanitize@url \@url }%
\providecommand \@url [1]{\endgroup\@href {#1}{\urlprefix }}%
\providecommand \urlprefix  [0]{URL }%
\providecommand \Eprint [0]{\href }%
\providecommand \doibase [0]{http://dx.doi.org/}%
\providecommand \selectlanguage [0]{\@gobble}%
\providecommand \bibinfo  [0]{\@secondoftwo}%
\providecommand \bibfield  [0]{\@secondoftwo}%
\providecommand \translation [1]{[#1]}%
\providecommand \BibitemOpen [0]{}%
\providecommand \bibitemStop [0]{}%
\providecommand \bibitemNoStop [0]{.\EOS\space}%
\providecommand \EOS [0]{\spacefactor3000\relax}%
\providecommand \BibitemShut  [1]{\csname bibitem#1\endcsname}%
\let\auto@bib@innerbib\@empty
\bibitem [{\citenamefont {Bardarson}\ \emph {et~al.}(2012)\citenamefont
  {Bardarson}, \citenamefont {Pollmann},\ and\ \citenamefont
  {Moore}}]{bardarson2012unbounded}%
  \BibitemOpen
  \bibfield  {author} {\bibinfo {author} {\bibfnamefont {J.~H.}\ \bibnamefont
  {Bardarson}}, \bibinfo {author} {\bibfnamefont {F.}~\bibnamefont {Pollmann}},
  \ and\ \bibinfo {author} {\bibfnamefont {J.~E.}\ \bibnamefont {Moore}},\
  }\href@noop {} {\bibfield  {journal} {\bibinfo  {journal} {Physical review
  letters}\ }\textbf {\bibinfo {volume} {109}},\ \bibinfo {pages} {017202}
  (\bibinfo {year} {2012})}\BibitemShut {NoStop}%
\bibitem [{\citenamefont {Serbyn}\ \emph {et~al.}(2013)\citenamefont {Serbyn},
  \citenamefont {Papi{\'c}},\ and\ \citenamefont
  {Abanin}}]{serbyn2013universal}%
  \BibitemOpen
  \bibfield  {author} {\bibinfo {author} {\bibfnamefont {M.}~\bibnamefont
  {Serbyn}}, \bibinfo {author} {\bibfnamefont {Z.}~\bibnamefont {Papi{\'c}}}, \
  and\ \bibinfo {author} {\bibfnamefont {D.~A.}\ \bibnamefont {Abanin}},\
  }\href@noop {} {\bibfield  {journal} {\bibinfo  {journal} {Physical review
  letters}\ }\textbf {\bibinfo {volume} {110}},\ \bibinfo {pages} {260601}
  (\bibinfo {year} {2013})}\BibitemShut {NoStop}%
\bibitem [{\citenamefont {Hartman}\ and\ \citenamefont
  {Maldacena}(2013)}]{Hartman:2013qma}%
  \BibitemOpen
  \bibfield  {author} {\bibinfo {author} {\bibfnamefont {T.}~\bibnamefont
  {Hartman}}\ and\ \bibinfo {author} {\bibfnamefont {J.}~\bibnamefont
  {Maldacena}},\ }\href {\doibase 10.1007/JHEP05(2013)014} {\bibfield
  {journal} {\bibinfo  {journal} {JHEP}\ }\textbf {\bibinfo {volume} {05}},\
  \bibinfo {pages} {014} (\bibinfo {year} {2013})},\ \Eprint
  {http://arxiv.org/abs/1303.1080} {arXiv:1303.1080 [hep-th]} \BibitemShut
  {NoStop}%
\bibitem [{\citenamefont {Liu}\ and\ \citenamefont
  {Suh}(2014{\natexlab{a}})}]{Liu:2013iza}%
  \BibitemOpen
  \bibfield  {author} {\bibinfo {author} {\bibfnamefont {H.}~\bibnamefont
  {Liu}}\ and\ \bibinfo {author} {\bibfnamefont {S.~J.}\ \bibnamefont {Suh}},\
  }\href {\doibase 10.1103/PhysRevLett.112.011601} {\bibfield  {journal}
  {\bibinfo  {journal} {Phys. Rev. Lett.}\ }\textbf {\bibinfo {volume} {112}},\
  \bibinfo {pages} {011601} (\bibinfo {year} {2014}{\natexlab{a}})},\ \Eprint
  {http://arxiv.org/abs/1305.7244} {arXiv:1305.7244 [hep-th]} \BibitemShut
  {NoStop}%
\bibitem [{\citenamefont {Liu}\ and\ \citenamefont
  {Suh}(2014{\natexlab{b}})}]{Liu:2013qca}%
  \BibitemOpen
  \bibfield  {author} {\bibinfo {author} {\bibfnamefont {H.}~\bibnamefont
  {Liu}}\ and\ \bibinfo {author} {\bibfnamefont {S.~J.}\ \bibnamefont {Suh}},\
  }\href {\doibase 10.1103/PhysRevD.89.066012} {\bibfield  {journal} {\bibinfo
  {journal} {Phys. Rev.}\ }\textbf {\bibinfo {volume} {D89}},\ \bibinfo {pages}
  {066012} (\bibinfo {year} {2014}{\natexlab{b}})},\ \Eprint
  {http://arxiv.org/abs/1311.1200} {arXiv:1311.1200 [hep-th]} \BibitemShut
  {NoStop}%
\bibitem [{\citenamefont {Kim}\ and\ \citenamefont {Huse}(2013)}]{Kim:2013bc}%
  \BibitemOpen
  \bibfield  {author} {\bibinfo {author} {\bibfnamefont {H.}~\bibnamefont
  {Kim}}\ and\ \bibinfo {author} {\bibfnamefont {D.~A.}\ \bibnamefont {Huse}},\
  }\href {\doibase 10.1103/PhysRevLett.111.127205} {\bibfield  {journal}
  {\bibinfo  {journal} {Phys. Rev. Lett.}\ }\textbf {\bibinfo {volume} {111}},\
  \bibinfo {pages} {127205} (\bibinfo {year} {2013})},\ \Eprint
  {http://arxiv.org/abs/1306.4306} {arXiv:1306.4306 [quant-ph]} \BibitemShut
  {NoStop}%
\bibitem [{\citenamefont {Kaufman}\ \emph {et~al.}(2016)\citenamefont
  {Kaufman}, \citenamefont {Tai}, \citenamefont {Lukin}, \citenamefont
  {Rispoli}, \citenamefont {Schittko}, \citenamefont {Preiss},\ and\
  \citenamefont {Greiner}}]{kaufman2016quantum}%
  \BibitemOpen
  \bibfield  {author} {\bibinfo {author} {\bibfnamefont {A.~M.}\ \bibnamefont
  {Kaufman}}, \bibinfo {author} {\bibfnamefont {M.~E.}\ \bibnamefont {Tai}},
  \bibinfo {author} {\bibfnamefont {A.}~\bibnamefont {Lukin}}, \bibinfo
  {author} {\bibfnamefont {M.}~\bibnamefont {Rispoli}}, \bibinfo {author}
  {\bibfnamefont {R.}~\bibnamefont {Schittko}}, \bibinfo {author}
  {\bibfnamefont {P.~M.}\ \bibnamefont {Preiss}}, \ and\ \bibinfo {author}
  {\bibfnamefont {M.}~\bibnamefont {Greiner}},\ }\href@noop {} {\bibfield
  {journal} {\bibinfo  {journal} {Science}\ }\textbf {\bibinfo {volume}
  {353}},\ \bibinfo {pages} {794} (\bibinfo {year} {2016})}\BibitemShut
  {NoStop}%
\bibitem [{\citenamefont {Hartman}\ and\ \citenamefont
  {Afkhami-Jeddi}(2015)}]{Hartman:2015apr}%
  \BibitemOpen
  \bibfield  {author} {\bibinfo {author} {\bibfnamefont {T.}~\bibnamefont
  {Hartman}}\ and\ \bibinfo {author} {\bibfnamefont {N.}~\bibnamefont
  {Afkhami-Jeddi}},\ }\href@noop {} {\  (\bibinfo {year} {2015})},\ \Eprint
  {http://arxiv.org/abs/1512.02695} {arXiv:1512.02695 [hep-th]} \BibitemShut
  {NoStop}%
\bibitem [{\citenamefont {Shenker}\ and\ \citenamefont
  {Stanford}(2014)}]{Shenker:2013pqa}%
  \BibitemOpen
  \bibfield  {author} {\bibinfo {author} {\bibfnamefont {S.~H.}\ \bibnamefont
  {Shenker}}\ and\ \bibinfo {author} {\bibfnamefont {D.}~\bibnamefont
  {Stanford}},\ }\href {\doibase 10.1007/JHEP03(2014)067} {\bibfield  {journal}
  {\bibinfo  {journal} {JHEP}\ }\textbf {\bibinfo {volume} {03}},\ \bibinfo
  {pages} {067} (\bibinfo {year} {2014})},\ \Eprint
  {http://arxiv.org/abs/1306.0622} {arXiv:1306.0622 [hep-th]} \BibitemShut
  {NoStop}%
\bibitem [{\citenamefont {Roberts}\ \emph {et~al.}(2015)\citenamefont
  {Roberts}, \citenamefont {Stanford},\ and\ \citenamefont
  {Susskind}}]{Roberts:2014isa}%
  \BibitemOpen
  \bibfield  {author} {\bibinfo {author} {\bibfnamefont {D.~A.}\ \bibnamefont
  {Roberts}}, \bibinfo {author} {\bibfnamefont {D.}~\bibnamefont {Stanford}}, \
  and\ \bibinfo {author} {\bibfnamefont {L.}~\bibnamefont {Susskind}},\ }\href
  {\doibase 10.1007/JHEP03(2015)051} {\bibfield  {journal} {\bibinfo  {journal}
  {JHEP}\ }\textbf {\bibinfo {volume} {03}},\ \bibinfo {pages} {051} (\bibinfo
  {year} {2015})},\ \Eprint {http://arxiv.org/abs/1409.8180} {arXiv:1409.8180
  [hep-th]} \BibitemShut {NoStop}%
\bibitem [{\citenamefont {Mezei}\ and\ \citenamefont
  {Stanford}(2017)}]{Mezei:2016wfz}%
  \BibitemOpen
  \bibfield  {author} {\bibinfo {author} {\bibfnamefont {M.}~\bibnamefont
  {Mezei}}\ and\ \bibinfo {author} {\bibfnamefont {D.}~\bibnamefont
  {Stanford}},\ }\href {\doibase 10.1007/JHEP05(2017)065} {\bibfield  {journal}
  {\bibinfo  {journal} {JHEP}\ }\textbf {\bibinfo {volume} {05}},\ \bibinfo
  {pages} {065} (\bibinfo {year} {2017})},\ \Eprint
  {http://arxiv.org/abs/1608.05101} {arXiv:1608.05101 [hep-th]} \BibitemShut
  {NoStop}%
\bibitem [{\citenamefont {Calabrese}\ and\ \citenamefont
  {Cardy}(2005)}]{Calabrese:2005in}%
  \BibitemOpen
  \bibfield  {author} {\bibinfo {author} {\bibfnamefont {P.}~\bibnamefont
  {Calabrese}}\ and\ \bibinfo {author} {\bibfnamefont {J.~L.}\ \bibnamefont
  {Cardy}},\ }\href {\doibase 10.1088/1742-5468/2005/04/P04010} {\bibfield
  {journal} {\bibinfo  {journal} {J. Stat. Mech.}\ }\textbf {\bibinfo {volume}
  {0504}},\ \bibinfo {pages} {P04010} (\bibinfo {year} {2005})},\ \Eprint
  {http://arxiv.org/abs/cond-mat/0503393} {arXiv:cond-mat/0503393 [cond-mat]}
  \BibitemShut {NoStop}%
\bibitem [{\citenamefont {Calabrese}\ and\ \citenamefont
  {Cardy}(2007)}]{Calabrese:2007rg}%
  \BibitemOpen
  \bibfield  {author} {\bibinfo {author} {\bibfnamefont {P.}~\bibnamefont
  {Calabrese}}\ and\ \bibinfo {author} {\bibfnamefont {J.}~\bibnamefont
  {Cardy}},\ }\href {\doibase 10.1088/1742-5468/2007/06/P06008} {\bibfield
  {journal} {\bibinfo  {journal} {J. Stat. Mech.}\ }\textbf {\bibinfo {volume}
  {0706}},\ \bibinfo {pages} {P06008} (\bibinfo {year} {2007})},\ \Eprint
  {http://arxiv.org/abs/0704.1880} {arXiv:0704.1880 [cond-mat.stat-mech]}
  \BibitemShut {NoStop}%
\bibitem [{\citenamefont {Casini}\ \emph {et~al.}(2016)\citenamefont {Casini},
  \citenamefont {Liu},\ and\ \citenamefont {Mezei}}]{Casini:2015zua}%
  \BibitemOpen
  \bibfield  {author} {\bibinfo {author} {\bibfnamefont {H.}~\bibnamefont
  {Casini}}, \bibinfo {author} {\bibfnamefont {H.}~\bibnamefont {Liu}}, \ and\
  \bibinfo {author} {\bibfnamefont {M.}~\bibnamefont {Mezei}},\ }\href
  {\doibase 10.1007/JHEP07(2016)077} {\bibfield  {journal} {\bibinfo  {journal}
  {JHEP}\ }\textbf {\bibinfo {volume} {07}},\ \bibinfo {pages} {077} (\bibinfo
  {year} {2016})},\ \Eprint {http://arxiv.org/abs/1509.05044} {arXiv:1509.05044
  [hep-th]} \BibitemShut {NoStop}%
\bibitem [{\citenamefont {Cotler}\ \emph {et~al.}(2016)\citenamefont {Cotler},
  \citenamefont {Hertzberg}, \citenamefont {Mezei},\ and\ \citenamefont
  {Mueller}}]{Cotler:2016acd}%
  \BibitemOpen
  \bibfield  {author} {\bibinfo {author} {\bibfnamefont {J.~S.}\ \bibnamefont
  {Cotler}}, \bibinfo {author} {\bibfnamefont {M.~P.}\ \bibnamefont
  {Hertzberg}}, \bibinfo {author} {\bibfnamefont {M.}~\bibnamefont {Mezei}}, \
  and\ \bibinfo {author} {\bibfnamefont {M.~T.}\ \bibnamefont {Mueller}},\
  }\href {\doibase 10.1007/JHEP11(2016)166} {\bibfield  {journal} {\bibinfo
  {journal} {JHEP}\ }\textbf {\bibinfo {volume} {11}},\ \bibinfo {pages} {166}
  (\bibinfo {year} {2016})},\ \Eprint {http://arxiv.org/abs/1609.00872}
  {arXiv:1609.00872 [hep-th]} \BibitemShut {NoStop}%
\bibitem [{\citenamefont {{Chandran}}\ and\ \citenamefont
  {{Laumann}}(2015)}]{2015PhRvB..92b4301C}%
  \BibitemOpen
  \bibfield  {author} {\bibinfo {author} {\bibfnamefont {A.}~\bibnamefont
  {{Chandran}}}\ and\ \bibinfo {author} {\bibfnamefont {C.~R.}\ \bibnamefont
  {{Laumann}}},\ }\href {\doibase 10.1103/PhysRevB.92.024301} {\bibfield
  {journal} {\bibinfo  {journal} {\prb}\ }\textbf {\bibinfo {volume} {92}},\
  \bibinfo {eid} {024301} (\bibinfo {year} {2015})},\ \Eprint
  {http://arxiv.org/abs/1501.01971} {arXiv:1501.01971 [cond-mat.dis-nn]}
  \BibitemShut {NoStop}%
\bibitem [{\citenamefont {Nahum}\ \emph {et~al.}(2016)\citenamefont {Nahum},
  \citenamefont {Ruhman}, \citenamefont {Vijay},\ and\ \citenamefont
  {Haah}}]{Nahum:2016muy}%
  \BibitemOpen
  \bibfield  {author} {\bibinfo {author} {\bibfnamefont {A.}~\bibnamefont
  {Nahum}}, \bibinfo {author} {\bibfnamefont {J.}~\bibnamefont {Ruhman}},
  \bibinfo {author} {\bibfnamefont {S.}~\bibnamefont {Vijay}}, \ and\ \bibinfo
  {author} {\bibfnamefont {J.}~\bibnamefont {Haah}},\ }\href@noop {} {\
  (\bibinfo {year} {2016})},\ \Eprint {http://arxiv.org/abs/1608.06950}
  {arXiv:1608.06950 [cond-mat.stat-mech]} \BibitemShut {NoStop}%
\bibitem [{\citenamefont {Nahum}\ \emph {et~al.}(2018)\citenamefont {Nahum},
  \citenamefont {Ruhman},\ and\ \citenamefont {Huse}}]{Nahum:2018pcr}%
  \BibitemOpen
  \bibfield  {author} {\bibinfo {author} {\bibfnamefont {A.}~\bibnamefont
  {Nahum}}, \bibinfo {author} {\bibfnamefont {J.}~\bibnamefont {Ruhman}}, \
  and\ \bibinfo {author} {\bibfnamefont {D.~A.}\ \bibnamefont {Huse}},\ }\href
  {\doibase 10.1103/PhysRevB.98.035118} {\bibfield  {journal} {\bibinfo
  {journal} {Phys. Rev.}\ }\textbf {\bibinfo {volume} {B98}},\ \bibinfo {pages}
  {035118} (\bibinfo {year} {2018})},\ \Eprint
  {http://arxiv.org/abs/1705.10364} {arXiv:1705.10364 [cond-mat.dis-nn]}
  \BibitemShut {NoStop}%
\bibitem [{\citenamefont {Ryu}\ and\ \citenamefont
  {Takayanagi}(2006{\natexlab{a}})}]{Ryu:2006bv}%
  \BibitemOpen
  \bibfield  {author} {\bibinfo {author} {\bibfnamefont {S.}~\bibnamefont
  {Ryu}}\ and\ \bibinfo {author} {\bibfnamefont {T.}~\bibnamefont
  {Takayanagi}},\ }\href {\doibase 10.1103/PhysRevLett.96.181602} {\bibfield
  {journal} {\bibinfo  {journal} {Phys. Rev. Lett.}\ }\textbf {\bibinfo
  {volume} {96}},\ \bibinfo {pages} {181602} (\bibinfo {year}
  {2006}{\natexlab{a}})},\ \Eprint {http://arxiv.org/abs/hep-th/0603001}
  {arXiv:hep-th/0603001 [hep-th]} \BibitemShut {NoStop}%
\bibitem [{\citenamefont {Ryu}\ and\ \citenamefont
  {Takayanagi}(2006{\natexlab{b}})}]{Ryu:2006ef}%
  \BibitemOpen
  \bibfield  {author} {\bibinfo {author} {\bibfnamefont {S.}~\bibnamefont
  {Ryu}}\ and\ \bibinfo {author} {\bibfnamefont {T.}~\bibnamefont
  {Takayanagi}},\ }\href {\doibase 10.1088/1126-6708/2006/08/045} {\bibfield
  {journal} {\bibinfo  {journal} {JHEP}\ }\textbf {\bibinfo {volume} {08}},\
  \bibinfo {pages} {045} (\bibinfo {year} {2006}{\natexlab{b}})},\ \Eprint
  {http://arxiv.org/abs/hep-th/0605073} {arXiv:hep-th/0605073 [hep-th]}
  \BibitemShut {NoStop}%
\bibitem [{\citenamefont {Hubeny}\ \emph {et~al.}(2007)\citenamefont {Hubeny},
  \citenamefont {Rangamani},\ and\ \citenamefont {Takayanagi}}]{Hubeny:2007xt}%
  \BibitemOpen
  \bibfield  {author} {\bibinfo {author} {\bibfnamefont {V.~E.}\ \bibnamefont
  {Hubeny}}, \bibinfo {author} {\bibfnamefont {M.}~\bibnamefont {Rangamani}}, \
  and\ \bibinfo {author} {\bibfnamefont {T.}~\bibnamefont {Takayanagi}},\
  }\href {\doibase 10.1088/1126-6708/2007/07/062} {\bibfield  {journal}
  {\bibinfo  {journal} {JHEP}\ }\textbf {\bibinfo {volume} {07}},\ \bibinfo
  {pages} {062} (\bibinfo {year} {2007})},\ \Eprint
  {http://arxiv.org/abs/0705.0016} {arXiv:0705.0016 [hep-th]} \BibitemShut
  {NoStop}%
\bibitem [{\citenamefont {Mezei}(2017)}]{Mezei:2016zxg}%
  \BibitemOpen
  \bibfield  {author} {\bibinfo {author} {\bibfnamefont {M.}~\bibnamefont
  {Mezei}},\ }\href {\doibase 10.1007/JHEP05(2017)064} {\bibfield  {journal}
  {\bibinfo  {journal} {JHEP}\ }\textbf {\bibinfo {volume} {05}},\ \bibinfo
  {pages} {064} (\bibinfo {year} {2017})},\ \Eprint
  {http://arxiv.org/abs/1612.00082} {arXiv:1612.00082 [hep-th]} \BibitemShut
  {NoStop}%
\bibitem [{\citenamefont {Mezei}\ and\ \citenamefont
  {Virrueta}(2019{\natexlab{a}})}]{Mezei:2019sla}%
  \BibitemOpen
  \bibfield  {author} {\bibinfo {author} {\bibfnamefont {M.}~\bibnamefont
  {Mezei}}\ and\ \bibinfo {author} {\bibfnamefont {J.}~\bibnamefont
  {Virrueta}},\ }\href@noop {} {\  (\bibinfo {year} {2019}{\natexlab{a}})},\
  \Eprint {http://arxiv.org/abs/1909.00919} {arXiv:1909.00919 [hep-th]}
  \BibitemShut {NoStop}%
\bibitem [{\citenamefont {Jonay}\ \emph {et~al.}(2018)\citenamefont {Jonay},
  \citenamefont {Huse},\ and\ \citenamefont {Nahum}}]{Jonay:2018yei}%
  \BibitemOpen
  \bibfield  {author} {\bibinfo {author} {\bibfnamefont {C.}~\bibnamefont
  {Jonay}}, \bibinfo {author} {\bibfnamefont {D.~A.}\ \bibnamefont {Huse}}, \
  and\ \bibinfo {author} {\bibfnamefont {A.}~\bibnamefont {Nahum}},\
  }\href@noop {} {\  (\bibinfo {year} {2018})},\ \Eprint
  {http://arxiv.org/abs/1803.00089} {arXiv:1803.00089 [cond-mat.stat-mech]}
  \BibitemShut {NoStop}%
\bibitem [{\citenamefont {Mezei}(2018)}]{Mezei:2018jco}%
  \BibitemOpen
  \bibfield  {author} {\bibinfo {author} {\bibfnamefont {M.}~\bibnamefont
  {Mezei}},\ }\href {\doibase 10.1103/PhysRevD.98.106025} {\bibfield  {journal}
  {\bibinfo  {journal} {Phys. Rev.}\ }\textbf {\bibinfo {volume} {D98}},\
  \bibinfo {pages} {106025} (\bibinfo {year} {2018})},\ \Eprint
  {http://arxiv.org/abs/1803.10244} {arXiv:1803.10244 [hep-th]} \BibitemShut
  {NoStop}%
\bibitem [{\citenamefont {Swingle}(2012{\natexlab{a}})}]{Swingle:2009bg}%
  \BibitemOpen
  \bibfield  {author} {\bibinfo {author} {\bibfnamefont {B.}~\bibnamefont
  {Swingle}},\ }\href {\doibase 10.1103/PhysRevD.86.065007} {\bibfield
  {journal} {\bibinfo  {journal} {Phys. Rev.}\ }\textbf {\bibinfo {volume}
  {D86}},\ \bibinfo {pages} {065007} (\bibinfo {year} {2012}{\natexlab{a}})},\
  \Eprint {http://arxiv.org/abs/0905.1317} {arXiv:0905.1317 [cond-mat.str-el]}
  \BibitemShut {NoStop}%
\bibitem [{\citenamefont {Swingle}(2012{\natexlab{b}})}]{Swingle:2012wq}%
  \BibitemOpen
  \bibfield  {author} {\bibinfo {author} {\bibfnamefont {B.}~\bibnamefont
  {Swingle}},\ }\href@noop {} {\  (\bibinfo {year} {2012}{\natexlab{b}})},\
  \Eprint {http://arxiv.org/abs/1209.3304} {arXiv:1209.3304 [hep-th]}
  \BibitemShut {NoStop}%
\bibitem [{\citenamefont {Pastawski}\ \emph {et~al.}(2015)\citenamefont
  {Pastawski}, \citenamefont {Yoshida}, \citenamefont {Harlow},\ and\
  \citenamefont {Preskill}}]{Pastawski:2015qua}%
  \BibitemOpen
  \bibfield  {author} {\bibinfo {author} {\bibfnamefont {F.}~\bibnamefont
  {Pastawski}}, \bibinfo {author} {\bibfnamefont {B.}~\bibnamefont {Yoshida}},
  \bibinfo {author} {\bibfnamefont {D.}~\bibnamefont {Harlow}}, \ and\ \bibinfo
  {author} {\bibfnamefont {J.}~\bibnamefont {Preskill}},\ }\href {\doibase
  10.1007/JHEP06(2015)149} {\bibfield  {journal} {\bibinfo  {journal} {JHEP}\
  }\textbf {\bibinfo {volume} {06}},\ \bibinfo {pages} {149} (\bibinfo {year}
  {2015})},\ \Eprint {http://arxiv.org/abs/1503.06237} {arXiv:1503.06237
  [hep-th]} \BibitemShut {NoStop}%
\bibitem [{\citenamefont {Hayden}\ \emph {et~al.}(2016)\citenamefont {Hayden},
  \citenamefont {Nezami}, \citenamefont {Qi}, \citenamefont {Thomas},
  \citenamefont {Walter},\ and\ \citenamefont {Yang}}]{Hayden:2016cfa}%
  \BibitemOpen
  \bibfield  {author} {\bibinfo {author} {\bibfnamefont {P.}~\bibnamefont
  {Hayden}}, \bibinfo {author} {\bibfnamefont {S.}~\bibnamefont {Nezami}},
  \bibinfo {author} {\bibfnamefont {X.-L.}\ \bibnamefont {Qi}}, \bibinfo
  {author} {\bibfnamefont {N.}~\bibnamefont {Thomas}}, \bibinfo {author}
  {\bibfnamefont {M.}~\bibnamefont {Walter}}, \ and\ \bibinfo {author}
  {\bibfnamefont {Z.}~\bibnamefont {Yang}},\ }\href {\doibase
  10.1007/JHEP11(2016)009} {\bibfield  {journal} {\bibinfo  {journal} {JHEP}\
  }\textbf {\bibinfo {volume} {11}},\ \bibinfo {pages} {009} (\bibinfo {year}
  {2016})},\ \Eprint {http://arxiv.org/abs/1601.01694} {arXiv:1601.01694
  [hep-th]} \BibitemShut {NoStop}%
\bibitem [{\citenamefont {Zhou}\ and\ \citenamefont
  {Nahum}(2019{\natexlab{a}})}]{Zhou:2018myl}%
  \BibitemOpen
  \bibfield  {author} {\bibinfo {author} {\bibfnamefont {T.}~\bibnamefont
  {Zhou}}\ and\ \bibinfo {author} {\bibfnamefont {A.}~\bibnamefont {Nahum}},\
  }\href {\doibase 10.1103/PhysRevB.99.174205} {\bibfield  {journal} {\bibinfo
  {journal} {Phys. Rev.}\ }\textbf {\bibinfo {volume} {B99}},\ \bibinfo {pages}
  {174205} (\bibinfo {year} {2019}{\natexlab{a}})},\ \Eprint
  {http://arxiv.org/abs/1804.09737} {arXiv:1804.09737 [cond-mat.stat-mech]}
  \BibitemShut {NoStop}%
\bibitem [{\citenamefont {Bao}\ and\ \citenamefont
  {Mezei}(2018)}]{Bao:2018wwd}%
  \BibitemOpen
  \bibfield  {author} {\bibinfo {author} {\bibfnamefont {N.}~\bibnamefont
  {Bao}}\ and\ \bibinfo {author} {\bibfnamefont {M.}~\bibnamefont {Mezei}},\
  }\href@noop {} {\  (\bibinfo {year} {2018})},\ \Eprint
  {http://arxiv.org/abs/1811.00019} {arXiv:1811.00019 [hep-th]} \BibitemShut
  {NoStop}%
\bibitem [{\citenamefont {Kudler-Flam}\ \emph {et~al.}(2019)\citenamefont
  {Kudler-Flam}, \citenamefont {Nozaki}, \citenamefont {Ryu},\ and\
  \citenamefont {Tan}}]{Kudler-Flam:2019wtv}%
  \BibitemOpen
  \bibfield  {author} {\bibinfo {author} {\bibfnamefont {J.}~\bibnamefont
  {Kudler-Flam}}, \bibinfo {author} {\bibfnamefont {M.}~\bibnamefont {Nozaki}},
  \bibinfo {author} {\bibfnamefont {S.}~\bibnamefont {Ryu}}, \ and\ \bibinfo
  {author} {\bibfnamefont {M.~T.}\ \bibnamefont {Tan}},\ }\href@noop {} {\
  (\bibinfo {year} {2019})},\ \Eprint {http://arxiv.org/abs/1906.07639}
  {arXiv:1906.07639 [hep-th]} \BibitemShut {NoStop}%
\bibitem [{\citenamefont {Agon}\ and\ \citenamefont
  {Mezei}(2019)}]{Agon:2019qgh}%
  \BibitemOpen
  \bibfield  {author} {\bibinfo {author} {\bibfnamefont {C.~A.}\ \bibnamefont
  {Agon}}\ and\ \bibinfo {author} {\bibfnamefont {M.}~\bibnamefont {Mezei}},\
  }\href@noop {} {\  (\bibinfo {year} {2019})},\ \Eprint
  {http://arxiv.org/abs/1910.12909} {arXiv:1910.12909 [hep-th]} \BibitemShut
  {NoStop}%
\bibitem [{\citenamefont {Mezei}\ and\ \citenamefont
  {Virrueta}(2019{\natexlab{b}})}]{Mezei:2019zyt}%
  \BibitemOpen
  \bibfield  {author} {\bibinfo {author} {\bibfnamefont {M.}~\bibnamefont
  {Mezei}}\ and\ \bibinfo {author} {\bibfnamefont {J.}~\bibnamefont
  {Virrueta}},\ }\href@noop {} {\  (\bibinfo {year} {2019}{\natexlab{b}})},\
  \Eprint {http://arxiv.org/abs/1912.11024} {arXiv:1912.11024 [hep-th]}
  \BibitemShut {NoStop}%
\bibitem [{\citenamefont {Zhou}\ and\ \citenamefont
  {Nahum}(2019{\natexlab{b}})}]{Zhou:2019pob}%
  \BibitemOpen
  \bibfield  {author} {\bibinfo {author} {\bibfnamefont {T.}~\bibnamefont
  {Zhou}}\ and\ \bibinfo {author} {\bibfnamefont {A.}~\bibnamefont {Nahum}},\
  }\href@noop {} {\  (\bibinfo {year} {2019}{\natexlab{b}})},\ \Eprint
  {http://arxiv.org/abs/1912.12311} {arXiv:1912.12311 [cond-mat.str-el]}
  \BibitemShut {NoStop}%
\bibitem [{Note1()}]{Note1}%
  \BibitemOpen
  \bibinfo {note} {In 2D every subregion is an interval, and the entropy grows
  linearly with time with the rate $v_E$. In 3D even spheres do not saturate
  the saturation time bound \cite {Mezei:2016zxg,Mezei:2018jco}, hence we chose
  to study the 4D case.}\BibitemShut {Stop}%
\bibitem [{\citenamefont {Leichenauer}(2014)}]{Leichenauer:2014nxa}%
  \BibitemOpen
  \bibfield  {author} {\bibinfo {author} {\bibfnamefont {S.}~\bibnamefont
  {Leichenauer}},\ }\href {\doibase 10.1103/PhysRevD.90.046009} {\bibfield
  {journal} {\bibinfo  {journal} {Phys. Rev.}\ }\textbf {\bibinfo {volume}
  {D90}},\ \bibinfo {pages} {046009} (\bibinfo {year} {2014})},\ \Eprint
  {http://arxiv.org/abs/1405.7365} {arXiv:1405.7365 [hep-th]} \BibitemShut
  {NoStop}%
\bibitem [{\citenamefont {Brakke}(1992)}]{brakke1992surface}%
  \BibitemOpen
  \bibfield  {author} {\bibinfo {author} {\bibfnamefont {K.~A.}\ \bibnamefont
  {Brakke}},\ }\href@noop {} {\bibfield  {journal} {\bibinfo  {journal}
  {Experimental mathematics}\ }\textbf {\bibinfo {volume} {1}},\ \bibinfo
  {pages} {141} (\bibinfo {year} {1992})}\BibitemShut {NoStop}%
\bibitem [{\citenamefont {Hayden}\ and\ \citenamefont
  {Preskill}(2007)}]{Hayden:2007cs}%
  \BibitemOpen
  \bibfield  {author} {\bibinfo {author} {\bibfnamefont {P.}~\bibnamefont
  {Hayden}}\ and\ \bibinfo {author} {\bibfnamefont {J.}~\bibnamefont
  {Preskill}},\ }\href {\doibase 10.1088/1126-6708/2007/09/120} {\bibfield
  {journal} {\bibinfo  {journal} {JHEP}\ }\textbf {\bibinfo {volume} {09}},\
  \bibinfo {pages} {120} (\bibinfo {year} {2007})},\ \Eprint
  {http://arxiv.org/abs/0708.4025} {arXiv:0708.4025 [hep-th]} \BibitemShut
  {NoStop}%
\bibitem [{\citenamefont {Sekino}\ and\ \citenamefont
  {Susskind}(2008)}]{Sekino:2008he}%
  \BibitemOpen
  \bibfield  {author} {\bibinfo {author} {\bibfnamefont {Y.}~\bibnamefont
  {Sekino}}\ and\ \bibinfo {author} {\bibfnamefont {L.}~\bibnamefont
  {Susskind}},\ }\href {\doibase 10.1088/1126-6708/2008/10/065} {\bibfield
  {journal} {\bibinfo  {journal} {JHEP}\ }\textbf {\bibinfo {volume} {10}},\
  \bibinfo {pages} {065} (\bibinfo {year} {2008})},\ \Eprint
  {http://arxiv.org/abs/0808.2096} {arXiv:0808.2096 [hep-th]} \BibitemShut
  {NoStop}%
\bibitem [{\citenamefont {Chesler}\ and\ \citenamefont
  {Yaffe}(2009)}]{Chesler:2008hg}%
  \BibitemOpen
  \bibfield  {author} {\bibinfo {author} {\bibfnamefont {P.~M.}\ \bibnamefont
  {Chesler}}\ and\ \bibinfo {author} {\bibfnamefont {L.~G.}\ \bibnamefont
  {Yaffe}},\ }\href {\doibase 10.1103/PhysRevLett.102.211601} {\bibfield
  {journal} {\bibinfo  {journal} {Phys. Rev. Lett.}\ }\textbf {\bibinfo
  {volume} {102}},\ \bibinfo {pages} {211601} (\bibinfo {year} {2009})},\
  \Eprint {http://arxiv.org/abs/0812.2053} {arXiv:0812.2053 [hep-th]}
  \BibitemShut {NoStop}%
\bibitem [{\citenamefont {Heller}\ \emph {et~al.}(2012)\citenamefont {Heller},
  \citenamefont {Janik},\ and\ \citenamefont {Witaszczyk}}]{Heller:2011ju}%
  \BibitemOpen
  \bibfield  {author} {\bibinfo {author} {\bibfnamefont {M.~P.}\ \bibnamefont
  {Heller}}, \bibinfo {author} {\bibfnamefont {R.~A.}\ \bibnamefont {Janik}}, \
  and\ \bibinfo {author} {\bibfnamefont {P.}~\bibnamefont {Witaszczyk}},\
  }\href {\doibase 10.1103/PhysRevLett.108.201602} {\bibfield  {journal}
  {\bibinfo  {journal} {Phys.\ Rev.\ Lett.}\ }\textbf {\bibinfo {volume}
  {108}},\ \bibinfo {pages} {201602} (\bibinfo {year} {2012})},\ \Eprint
  {http://arxiv.org/abs/1103.3452} {arXiv:1103.3452 [hep-th]} \BibitemShut
  {NoStop}%
\bibitem [{\citenamefont {Kovtun}\ \emph {et~al.}(2003)\citenamefont {Kovtun},
  \citenamefont {Son},\ and\ \citenamefont {Starinets}}]{Kovtun:2003wp}%
  \BibitemOpen
  \bibfield  {author} {\bibinfo {author} {\bibfnamefont {P.}~\bibnamefont
  {Kovtun}}, \bibinfo {author} {\bibfnamefont {D.~T.}\ \bibnamefont {Son}}, \
  and\ \bibinfo {author} {\bibfnamefont {A.~O.}\ \bibnamefont {Starinets}},\
  }\href {\doibase 10.1088/1126-6708/2003/10/064} {\bibfield  {journal}
  {\bibinfo  {journal} {JHEP}\ }\textbf {\bibinfo {volume} {10}},\ \bibinfo
  {pages} {064} (\bibinfo {year} {2003})},\ \Eprint
  {http://arxiv.org/abs/hep-th/0309213} {arXiv:hep-th/0309213} \BibitemShut
  {NoStop}%
\end{thebibliography}%
\clearpage

\section{Supplemental Material}

\subsection{Spheres in charged quenches}

\begin{center}
\begin{figure}[!h]
\includegraphics[scale=0.6]{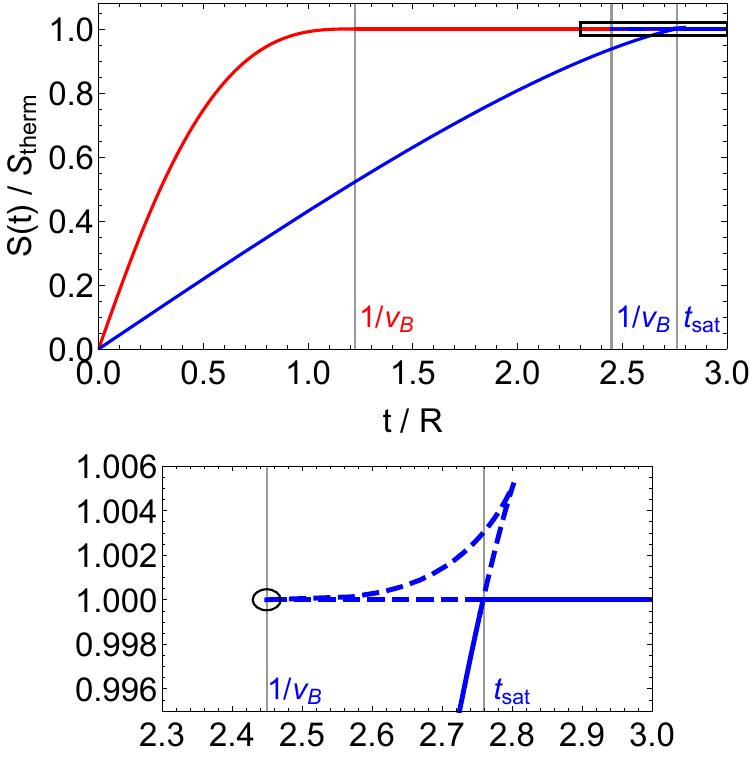}
\caption{ On the top figure we are including the time evolution of the entropy in a neutral quench (red) and charged quench with $q=0.62$ (blue). On the bottom figure we zoom in onto the near saturation behavior in the charged quench, and plot the analytically obtainable merging favored (minimal) and disfavored (non-minimal) branches of timelike membranes. The $v_B$ light sheet is indicated by a black circle and lies on the disfavored branch. There is a third branch of horizontal (spacelike) membranes that are responsible for the saturation of entropy. 
\label{fig:Sphere}}
\end{figure}
\end{center}

In the main text we explained that there is another convincing example for the universality of the scenario that makes the  $v_B$ light sheet membrane not globally minimal: the analytically solvable case of a spherical $A$ for the membrane tension describing a charged quench  that equilibrates to the grand canonical ensemble with nonzero chemical potential, plotted in Fig.~\ref{fig:membrane} (right). This problem was already solved in \cite{Mezei:2016zxg}, and we are including it in the SM to make the paper self-contained. The time evolution of the entropy is plotted in Fig.~\ref{fig:Sphere}. The neutral quench is dual to the formation of a Schwarzschild black brane that is expected to be fast at all tasks, while the charged quench for large chemical potential creates a near extremal charged black brane that is expected to be slower at certain tasks \cite{Leichenauer:2014nxa}.

\subsection{Results for ellipses and ellipsoids}

\begin{center}
\begin{figure}[!htbp]
\includegraphics[width=0.99\linewidth]{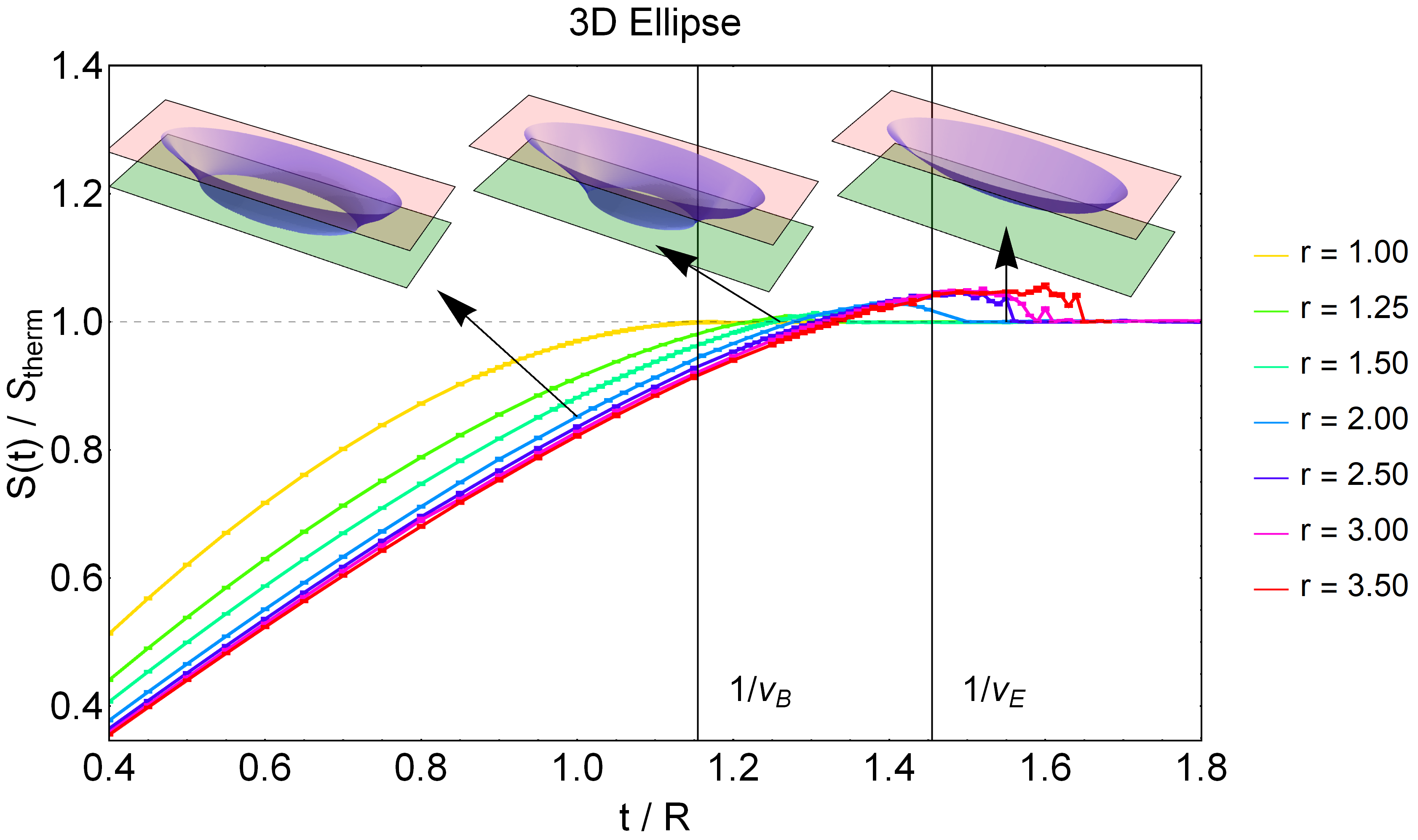}
\caption{ 
We show the entanglement entropy of ellipse subregions of axis ratios $r$ as a function of time $t/R$, where $2 R$ is length of the short axis. The entanglement entropy saturates when $S[A(t)]$ equals the thermal value, even though as shown there exist membranes  that are not globally minimal. We show three membranes at representative times for the ratio $r=3$.
\label{fig:ellipse3d}}
\end{figure}
\end{center}

In Fig.~\ref{fig:ellipse3d} we show representative examples in 3D for a neutral quench, where the entangling region is an ellipse. While from the plot it looks like  the disk saturates entanglement entropy at $t_B$, in reality $t_\text{sat}>t_B$ by an extremely small amount as was explained in detail in \cite{Mezei:2016zxg,Mezei:2018jco} and mentioned in the main text. For other ellipses shown $t_\text{sat}$ is bigger than $t_B$ by a visible amount.

\begin{center}
\begin{figure}[!htbp]
\includegraphics[width=0.99\linewidth]{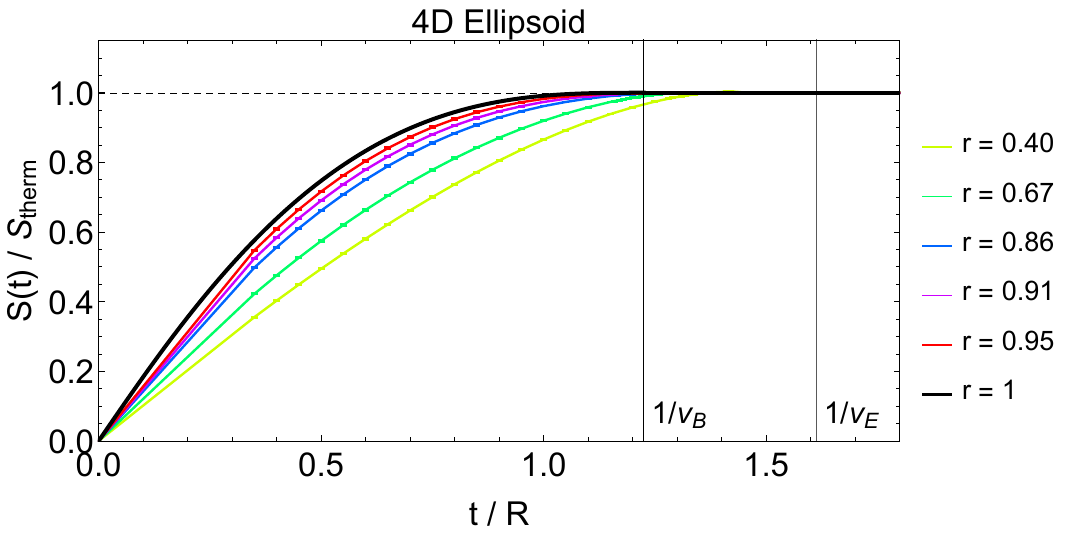}
\includegraphics[width=0.99\linewidth]{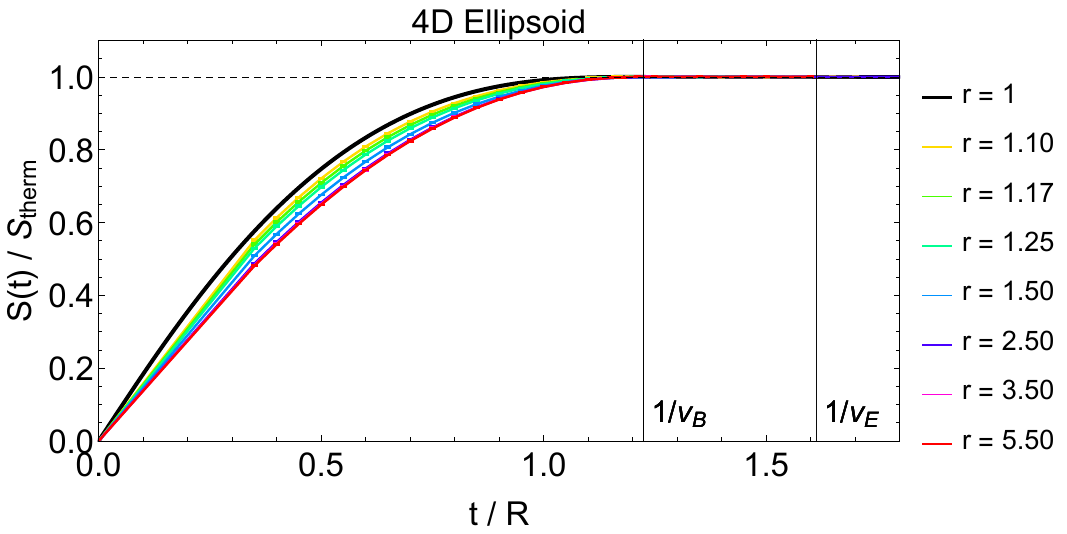}
\caption{ 
We show the entanglement entropy of regions bounded by ellipsoids of axis ratios $r$ as a function of time $t/R$, where $2 R$ is length of the short axis. $r<1$ for oblate and $r>1$ for prolate ellipsoids, which is consistent with the choice we made for stadia on Figs.~\ref{fig:examples} and~\ref{fig:examplesR}. Since the entanglement entropy smoothly approaches its thermal value it is hard to define an exact saturation time, but our results are consistent with all surfaces saturating at $t_{\rm sat} =t_B$ for $r>1$. The results are qualitatively the same as for the stadia presented in Fig.~\ref{fig:manystadia} in the main text.
\label{fig:ellipse4d}}
\end{figure}
\end{center}

In Fig.~\ref{fig:ellipse4d} we show the analog figure to Fig.~\ref{fig:manystadia} for stadia as presented in the main text. Again we see a saturation at the butterfly time $t_B$ for elongated ellipses (rotated along the long axis), while for squashed ellipses (rotated along the short axis) we have $t_\text{sat}>t_B$, in complete analogy with the stadia. Numerically it is however not easy to get a $t_\text{sat}$ as large as observed in Fig.~\ref{fig:manystadia}, which could indicate that for ellipsoids the maximum saturation time is less than $t_E$. (Note that in this case it is not possible to analytically compute the $r\rightarrow 0$ or $r\rightarrow \infty$ limits.)

\pagebreak

\subsection{Numerical methods}

The instinct of a physicist when seeing an action like \eqref{AreaFunctScaled} is to write down the Euler-Lagrange PDE and solve it numerically. For our problem this route runs into hurdles due to the typical cusp formation on the membrane. We found it easier to minimize the functional  \eqref{AreaFunctScaled} directly using Surface Evolver. To this end, it is advantageous to change from Minkowski to Euclidean signature. This amounts to changing:
\es{MinkEuc}{
{d\text{area}_\text{Mink}\ov \sqrt{1-v^2}} \to {d\text{area}_\text{Euc}\ov \sqrt{1+v^2}}\,,
}
with the expression $v^2=n_t^2/(1+n_t^2)$ unchanged between the two signatures.

\begin{figure*}
\includegraphics[width=0.32\linewidth]{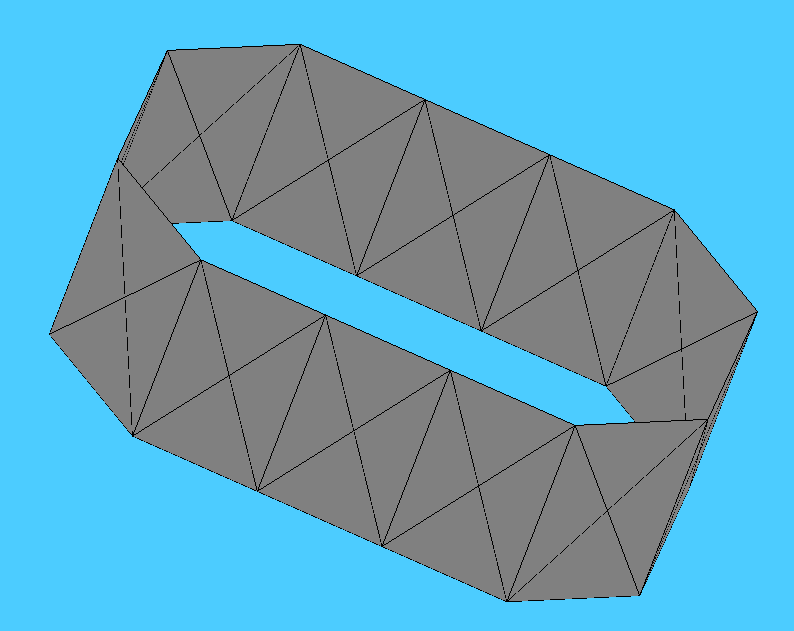}
\includegraphics[width=0.32\linewidth]{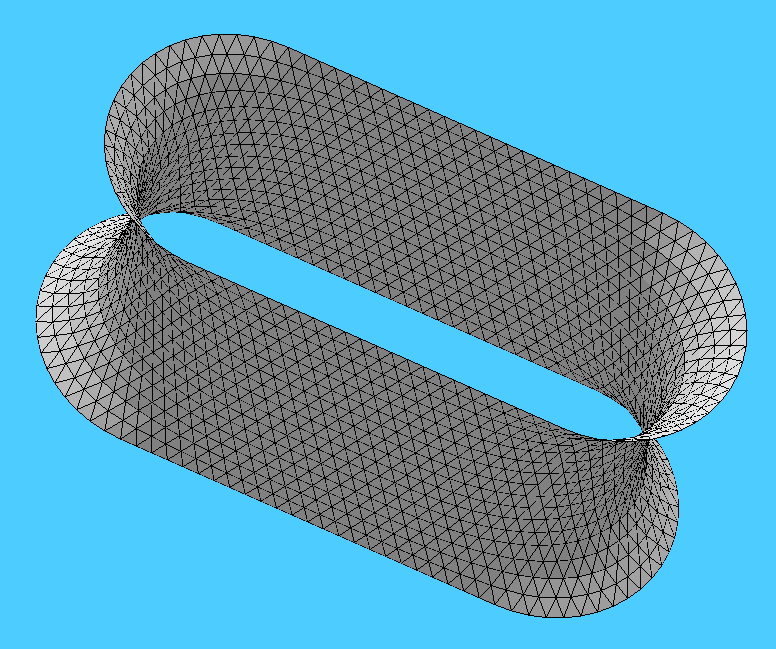}
\includegraphics[width=0.32\linewidth]{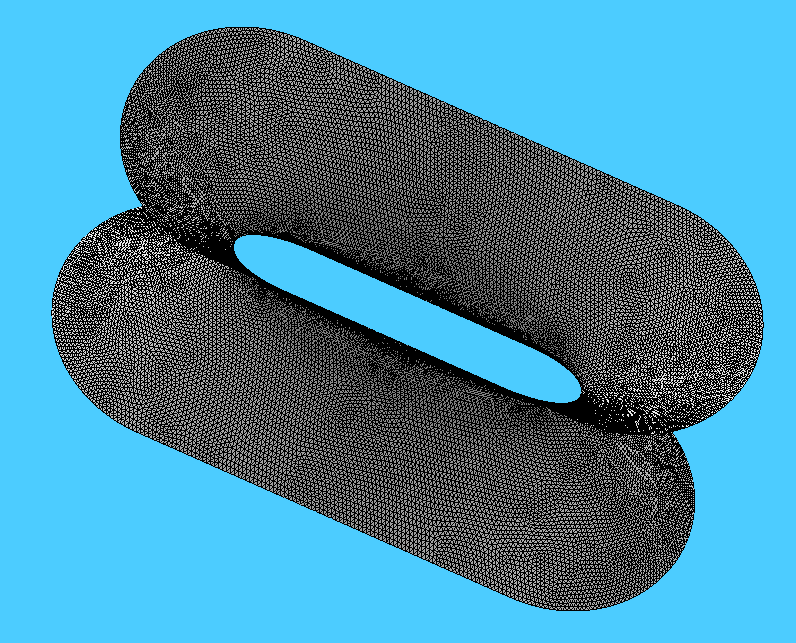}
\caption{ 
Initial condition for the stadium at the roughest level of triangulation (left), together with an intermediate result (middle) and the final shape (right). Even though there is an 8-fold symmetry we evolve the full surface and use the difference in energy of the eight regions as an estimate of our numerical uncertainty.
\label{fig:stadiumini}}
\end{figure*}

We present our numerical procedure for the stadia in 4D, which is hardest to converge; the other surfaces are obtained using similar procedures. We implemented the boundary conditions by doubling the system, and anchoring the membrane on $A$ and its mirror image on the resulting two slabs. Half of the resulting minimal membrane will be perpendicular to the $t=0$ surface by symmetry. For the Surface Evolver we initialize the surface using eight segments (four half-circles and four lines) each having three points on them, as displayed in Fig. \ref{fig:stadiumini} (left). The lower segments are located at $z=x_3=0$, whereas the top ones are at $z=2t/R$. Note that the Surface Evolver upon refinement automatically places new points on the relevant half-circle.

For the energy functional Surface Evolver labels the normal vector as $(x4, x5, x6)$, where $x6$ is in the timelike direction labelled $\hat{t}^\mu$ in the main text. This means the energy functional reads $\frac{\pi  \left| \text{x2}\right| }{\sqrt[4]{1-v^2} \sqrt{v^2+1}}$, where $v=\frac{\text{x6}}{\sqrt{\text{x4}^2+\text{x5}^2+\text{x6}^2}}$ and $x2$ is the Jacobian from our imposed rotational symmetry. 

Besides changing to Euclidean signature, we have to implement another modification for better numerical stability. The membrane tension function ${\cal E}(v)$ diverges for $\abs{v}\to 1$. As mentioned in the main text one can show that minimal membranes obey $\abs{v}\leq v_B<1$, hence we are allowed to modify ${\cal E}(v)$ for $v>v_B$ at will.
In practice we use the following energy functional, multiplied by the area element $\sqrt{x4^2+x5^2+x6^2}$:
\begin{lstlisting}
quantity esurface energy method 
facet_general_integral
scalar_integrand: 
1.*((x6^2/(x4^2+x5^2+10^(-15.0))<=0.8)?
(((x4^2 + x5^2+10^(-15.0))^0.75*abs(x2))/
(abs(x4^2 + x5^2 - x6^2)+10^(-15.0))^0.25)
:
((((5*5^0.25 - 4*a)*x4^2 +
 (5*5^0.25 - 4*a)*x5^2 + 5*a*x6^2)*abs(x1))/
 (5.*sqrt(x4^2 + x5^2+10^(-15.0))))
),
\end{lstlisting}
where the if statement with $>0.8$ is not realized in physical solutions. In practice one needs this second term (we use $a=10$) during the relaxation, where part of the surface may (temporarily) surpass the bound $\abs{v}\leq v_B$, which could (and often would) lead to division by zero errors.

After initialization we then run the following relaxation scheme:
\begin{lstlisting}
gVu  := {V; u; g 25;}
gogo := {
  scale:=0.001;
  r 2; gVu 50;
  scale:=0.0025;
  r; g 12000;
  V; u; g 29000;
  gVu 8000;
  scale:=0.0005;
  g 30000;
  r; gVu 15;
  g 20000;
  r; gg  10; 
  g 100000
},
\end{lstlisting}
where $V$ averages vertices using the centroids of adjacent facets, $u$ is called equiangulation and switches triangles on a face if this leads to a more equiangular triangulization. $g$ is the main relaxing routine, which is using `scale' as a multiplication factor. Of course a higher scale means faster relaxation, but it can also increase numerical artifacts, so that we found a final scale of 0.0005 to be optimal. The final g 100000 takes most time (of order several hours), but it can be reduced to get accurate results for most points (the most difficult ones are the shapes close to saturation, where the surface is almost minimal already).

\section{Selected comments on the main text}

We note that for times $t\lesssim t_\text{loc}$, in time reflection symmetric quench states, the entropy grows quadratically \cite{Liu:2013iza,Liu:2013qca,Mezei:2019sla}.  In the hydrodynamic limit, we are only describing regions and times for which $R,t\gg t_\text{loc}$
 (with $t/R$ fixed), and the leading extensive piece of the entropy. The entropy produced during quadratic growth is subleading in the hydrodynamic limit, hence it is not captured by the current membrane theory study. 
 
 In the main text we only considered membranes that describe the growth of entropy, but not its saturation. Besides the membranes discussed thus far, we also have to allow for spacelike membranes that are horizontal (formally with $v=\infty$), for which the action is equal to the area. When they have minimal area, these describe the saturation of entropy. We also allow for membranes that have horizontal and timelike parts, but they do not play a role in our discussion.
 
We now comment on a possible improvement on the upper bound on the entropy \eqref{Sbound}. It was claimed in \cite{Jonay:2018yei,Mezei:2018jco} that this membrane is the minimal membrane of 
the maximal membrane tension function consistent with the constraints explained below \eqref{AreaFunctScaled}: ${\cal E}_\text{max}\le(v\ri)=v_E+\le(1-{v_E\ov v_B}\ri)\abs{v}$. This is imprecise: the membranes in the theory  ${\cal E}_\text{max}\le(v\ri)$ are obtained by matching the light sheet and vertical tube parts along any spacelike hypersurface, not necessarily along a constant time slice, and hence generically have smaller action. A bit of thought suggests that the bound of  \cite{Mezei:2016wfz} in relativistic theories can be tightened  by sewing together the two bounds used in the derivation along  a matching spacelike hypersurface to reproduce the entropy determined by  ${\cal E}_\text{max}\le(v\ri)$. Since this refinement makes analytic progress much harder, we do not pursue it in this work.

\end{document}